\documentclass[pdflatex,sn-mathphys]{sn-jnl}% Math and Physical Sciences Reference Style
\usepackage{cleveref}
\jyear{2021}%

\theoremstyle{thmstyleone}%
\theoremstyle{thmstyletwo}%

\theoremstyle{thmstylethree}%

\begin{document}

\title[Deep learning-assisted localisation of nanoparticles]{Deep Learning-Assisted Localisation of Nanoparticles in synthetically generated two-photon microscopy images}

\author*[1]{\fnm{} \sur{Rasmus Netterstrøm}}\email{rane@di.ku.dk}
\equalcont{These authors contributed equally to this work.}

\author[2]{\fnm{} \sur{Nikolay Kutuzov}}\email{nkutuzov@sund.ku.dk}
\equalcont{These authors contributed equally to this work.}

\author[1]{\fnm{} \sur{Sune Darkner}}\email{darkner@di.ku.dk}

\author[1]{\fnm{} \sur{Maurits Jørring Pallesen}}\email{mauritsjpallesen@gmail.com}

\author[2,3]{\fnm{} \sur{Martin Johannes Lauritzen}}\email{mlauritz@sund.ku.dk}

\author[1]{\fnm{} \sur{Kenny Erleben}}\email{kenny@di.ku.dk}

\author[1]{\fnm{} \sur{Fran\c{c}ois Lauze}}\email{francois@di.ku.dk}

\affil*[1]{\orgdiv{Department of computer science}, \orgname{University of Copenhagen}, \orgaddress{\street{Universitetsparken}, \city{København}, \postcode{2100}, \country{Denmark}}}

\affil[2]{\orgdiv{Department of Neuroscience}, \orgname{University of Copenhagen}, \orgaddress{\street{Blegdamsvej}, \city{København}, \postcode{2200}, \country{Denmark}}}

\affil[3]{\orgdiv{Department of Clinical Neurophysiology}, \orgname{Rigshospitalet}, \orgaddress{\street{Blegdamsvej}, \city{København}, \postcode{2100}, \country{Denmark}}}

\abstract{
Tracking single molecules is instrumental for quantifying the transport of molecules and nanoparticles in biological samples, e.g., in brain drug delivery studies. Existing intensity-based localisation methods are not developed for imaging with a scanning microscope, typically used for in vivo imaging. 
Low signal-to-noise ratios, movement of molecules out-of-focus, and high motion blur on images recorded with scanning two-photon microscopy (2PM) in vivo pose a challenge to the accurate localisation of molecules. Using data-driven models is challenging due to low data volumes, typical for in vivo experiments. We developed a 2PM image simulator to supplement scarce training data.  
The simulator mimics realistic motion blur, background fluorescence, and shot noise observed in vivo imaging. 
Training a data-driven model with simulated data improves localisation quality in simulated images and shows why intensity-based methods fail.  
}



\maketitle

\section{Introduction}\label{sec1}

Tracking single molecules or nanoparticles helps to solve a variety of biomedical problems, e.g., drug design and delivery \cite{Kucharz2021}, studying the regulation of gene expression \cite{Elf2007}, determining mechanisms of viral infections \cite{Liu2012}, and many others. Single-molecule (or particle) tracking (SMT) is becoming a popular tool to study mechanisms of transport at the nanoscale in living cells and tissues \cite{Godin2017}. In vivo imaging, especially in scattering tissues like the brain, becomes an ultimate challenge for SMT due to (i) low photon budgets causing low signal-to-noise ratio on recorded images, (ii) motion artefacts caused by the animal’s breathing, heartbeat, movement, and (iii) severe motion blur caused by the slowness of scanning microscopy typically used in vivo. These factors challenge the accurate localisation and tracking of molecules on images recorded in vivo, for example with two-photon microscopy (2PM; Fig.~\ref{fig:2pm}\textbf{a}). 

Object localisation and detection is a core topic in computer vision. Feature-based detection, using SIFT \cite{SIFT}, HoG \cite{dalal_triggs:2005}, to cite the most classically used, were replaced by deep learning (DL) approaches, notably \cite{ren-etal:2015,yolo1} and many were built for specialised situations in medical imaging. On images of single molecules recorded with CCD or EMCCD cameras, localisation is achievable even for images with low signal-to-noise ratio or(and) moderate motion blur \cite{Mortensen2010,Mortensen2021}. The localisation algorithms, however, require writing down an expression for the distribution of fluorescence in recorded images, which may not be possible for imaging moving molecules with 2PM: random motion of molecules overlaid with the laser beam’s motion (Fig.~\ref{fig:2pm}\textbf{b}) can make modelling the distribution of fluorescence on a recorded image prohibitively complex. Neural networks have the potential to approximate these complex distributions of fluorescence and return the desired position of the molecule. Deep learning (DL) has been predominantly used for the localisation of immobile molecules with wide-field microscopes equipped with a CCD or EMCCD camera \cite{Petrov2017,Boyd2018,Nehme2018} in various in vitro experiments but not yet for localisation of moving molecules with 2PM in living animals.

Experimental data recorded in vivo can be so scarce and expensive to collect that training of the neural networks is more likely to rely on artificial data. Simulating large training data sets, informed by well-known theories of microscope image formation, has been used to train deep networks to localise immobile objects on images recorded with a CCD or EMCCD camera \cite{Petrov2017,Boyd2018}. In vivo imaging in scattering tissues, e.g. in the brain, however, is typically done with scanning 2PM, which records images pixel-by-pixel, line-by-line (Fig.~\ref{fig:2pm}\textbf{b}), as opposed to collecting all image pixels simultaneously with a CCD camera. 
Here we show how to train a neural network to localise single simulated quantum dots (QDs) and how the commonly used intensity-based detections fails to localise the QD on pixel-by-pixel and line-by-line simulated images. We used a combination of realistic simulated data, based on modelling diffusion of QDs during exposure and realistic motion of the laser beam. Many DL approaches have been used for the detection and localisation of objects, both with natural images and medical imaging modalities too.  We demonstrate how a neural network used for object localisation and detection for natural images,  can accommodate the QD detection and localisation task by  training it on simulated data, incorporating effects of motion blur caused by very different rates between diffusion motion and camera motion, and then allow localisation of QDs on images sampled from a simulated test dataset.

\section{Results}\label{sec2}
%\subsection{Data description}
 %We simulated a training dataset of 29838 images and three testing datasets of 2570, 2410 and 1870 images. All images were 24 pixels tall and 25 pixels wide, with a set of parameters reproducing in vivo experiments. Trajectories of the QDs were simulated with a set of diffusion coefficients between $0.75 \mu m^2/s$ and $1.75 \mu m^2/s$. All images contain 1 QD. The pixel size was $0.07 \mu m$, the pixel exposure time was $2 \mu s$ and the retracing time was $1.1ms$. 

%\subsection{Deep Learning Model}
%We use the PyTorch implementation of FRCNN-keypoint with a resNet backbone \cite{resnet} pre-trained on the MS COCO dataset \cite{coco}.

%\subsection{Train Configuration}
%To train our model, we used: 20 epochs, a training batch size of 6 images, a validation batch size of 4 images, and a stochastic gradient descent (SGD) with a learning rate of 0.001. Each epoch resulted in one model; we picked the model with the highest validation accuracy. 

%\subsection{Results}

\begin{figure}[h!]
    \centering
        \includegraphics[width=\textwidth]{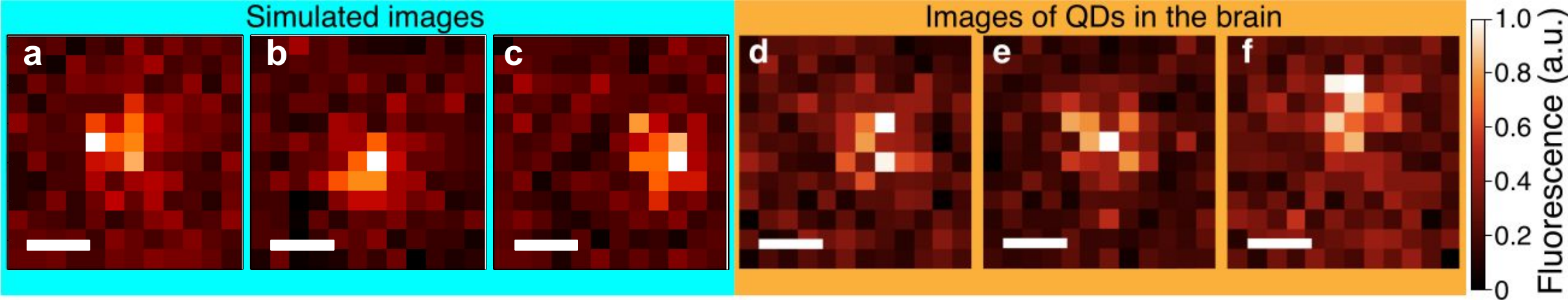} 
    \caption{\textbf{Comparison of three artificially simulated images (\textbf{a-c}) with real experimental images of QDs in the brain (\textbf{d-f}), recorded with 2PM.}
    Note how the simulated data reproduces the motion blur: panels \textbf{a}, \textbf{b}, and \textbf{f}, show elongated fluorescence intensity distributions in the vertical direction due to the movement of the QDs.
    Scale bars are 500 nm wide.
    }
    \label{fig:compare}
\end{figure}

We started by developing a protocol to generate realistic synthetic 2PM images for training the FRCNN model \cite{maskfcnn,pytorch-keypoint-rcnn, pytorch}, later used for the localization of QDs.
By combining a simple Brownian-like motion of QDs with a deterministic motion of the laser beam we reproduced motion blur characteristics for scanning microscopy (Fig.~\ref{fig:compare}).
We experimentally measured point spread function (PSF) dimensions and estimated fluorescence background from experimental images ($N_{bkg}$ in Eq.~\ref{eq:psf}) to mimic real images as close as possible.
Finally, with Poisson noise, we simulated images that looked very similar to images recorded in the brain in vivo (Fig.~\ref{fig:compare}).
%Note that the generation of these synthetic images is very simple computationally, which allows generating large amounts of realistic training data very fast.

\begin{figure}[h!]
    \centering
        \includegraphics[width=\textwidth]{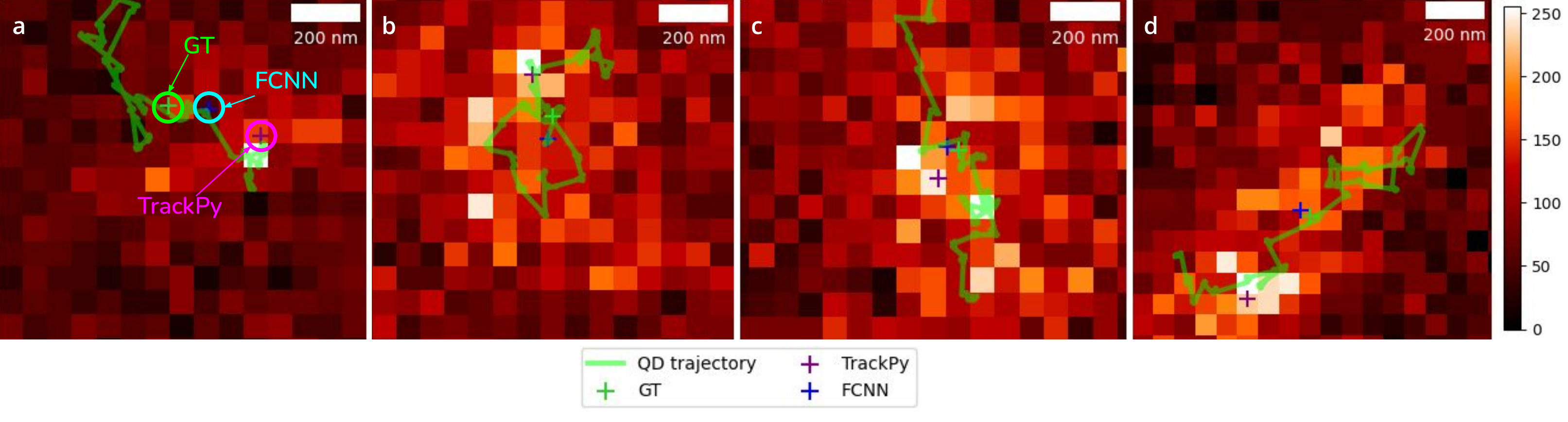} 
    \caption{\textbf{Localization of QDs on simulated images (\textbf{a}-\textbf{d})}
    Blue and purple crosses show the localisation of QDs estimated with FCNN and TrackPy, respectively. Green curves show simulated trajectories; green crosses show trajectory centroids.
    }
    \label{fig:localisation}
\end{figure}

In \Cref{fig:localisation}, we highlight the primary issue of intensity-based detectors in four images. The intensity-based detector, TrackPy \cite{trackpy}, detects high-intensity pixels even though the high-intensity pixels are not the trajectory centroid of the QD. Training directly on the centroid coordinate enables FCNN to detect the trajectory centroid of the QD with higher accuracy. 
Then we conducted three experiments to demonstrate further the value of including the trajectory centroid in the DL training pipeline. Each experiment had a different diffusion coefficient that provided a different motion blur. The first experiment had a diffusion coefficient of $0.75\mu m^2/s$, the second of $1 \mu m^2 / s$ and the third of $1.75 \mu m^2 / s$. Besides the different diffusion coefficients, we tested the robustness of each model by changing the epsilon-threshold ($\varepsilon$) of the precision-recall curve; The epsilons were $1$,$2$ and $3$, corresponding to a distance of $70nm$, $140nm$ and $210nm$ given a pixel size of $0.07 \mu m$. 

\begin{figure}[h!]
    \centering
        \includegraphics[width=\textwidth]{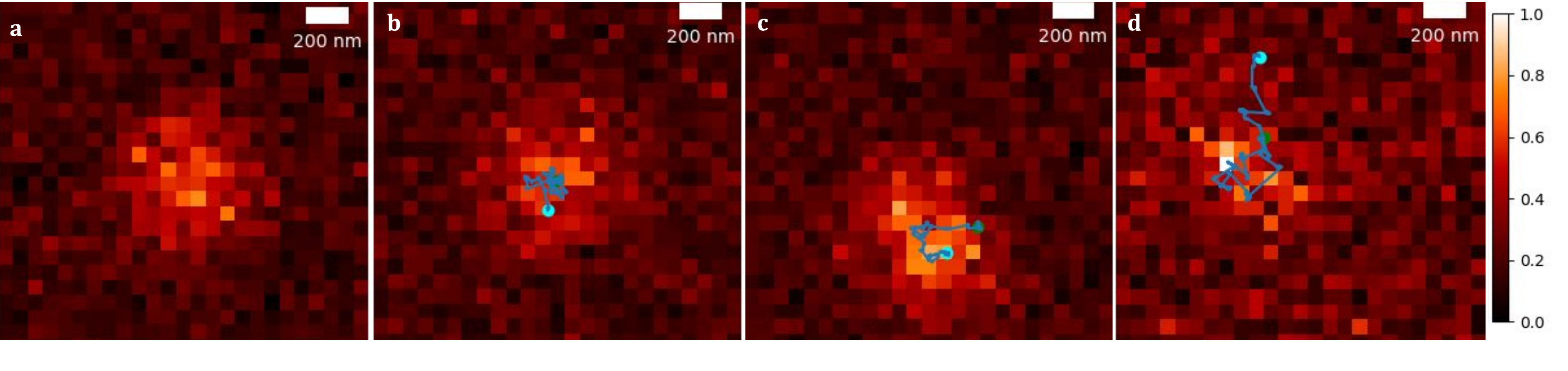} 
    \caption{\textbf{Four artificially simulated images (a-d)} with diffusion coefficient going from 0 to $1.75 \mu m^2 / s$. In images b-d, the green circle is the starting point of the trajectory, and the cyan circle is the ending point. (a) show a QD that diffuses with a diffusion coefficient of $0\mu m^2/s$, showing the microscope PSF with noise. (b-d) show QD diffusing with increasing diffusion coefficient: $0.75, 1 \text{ and } 1.75 \mu m^2/s$.     }
    \label{fig:testdata}
\end{figure}

\begin{table}[h]
\begin{center}
\begin{minipage}{\textwidth}
\caption{This table shows the AUC-score of all three experiments. In the first column, the QD diffuse with $0.75 \mu m^2/s$, in the second column, the QD diffuse with $1 \mu m^2/s$ and in the third column, the QD diffuse with $1.75 \mu m^2/s$. In all three columns, the FCNN model achieves the best AUC-score. }\label{tab:res:1}\label{tab:auc}
\begin{tabular*}{\textwidth}{@{\extracolsep{\fill}}lccccccccc@{\extracolsep{\fill}}}
\toprule%
D & \multicolumn{3}{@{}c@{}}{$0.75\mu m^2/s$} & \multicolumn{3}{@{}c@{}}{$1\mu m^2 /s$}  & \multicolumn{3}{@{}c@{}}{$1.75 \mu m^2 /s$}\\\cmidrule{2-4}\cmidrule{5-7}\cmidrule{8-10}%
$nm$ & $70$ & $140$ & $210$ & $70$ & $140$ & $210$ & $70$ & $140$ & $210$\\
\midrule
TrackPy   & 0.06 & 0.42 & 0.73 & 0.04 & 0.30 & 0.61 &  0.02 & 0.19 & 0.47\\
FCNN    & \textbf{0.48} & \textbf{0.91} & \textbf{0.99}  &  \textbf{0.41} & \textbf{0.87} & \textbf{0.98} & \textbf{0.26} & \textbf{0.72 }& \textbf{0.92} \\
\botrule
\end{tabular*}
\end{minipage}
\end{center}
\end{table}

\subsubsection{Experiment 1 - $0.75\;\mu m^2/s$}
% need 2000 images with diffusion 0.75 mu m^2/s
% testing threshold from 3-1 mu m
This experiment concentrates on localising the trajectory centroid of a QD when 
point sources of light closely resemble the microscope’s PSF. That ensures the trajectory centroid is close to the high-intensity pixels, hence less motion blur. We choose a realistic diffusion coefficient value of $0.75\mu m^2 / s$. \textbf{\Cref{fig:testdata} (b)} shows a QD moving with a diffusion coefficient of $0.75\mu m^2 / s$, allowing less motion blur.  
\textbf{\Cref{tab:auc}, column 1} demonstrates that TrackPy and FCNN localise the trajectory centroid for localisation within $210nm$ with an AUC score of $0.73$ and $0.99$.  

\subsubsection{Experiment 2 - $1\;\mu m^2/s$}
% need 2000 images with diffusion 1 mu m^2/s
% testing threshold from 3-1 mu m
We increase the diffusion coefficient, thus allowing more motion blur. \textbf{\Cref{fig:testdata} (c)} illustrates how the trajectory steps get larger, and the high-intensity pixels diverge from the microscope PSF; the distance between the trajectory centroid and high-intensity pixels becomes unpredictable. \textbf{\Cref{tab:auc}, column 2} tells the immediate effect of more motion blur in the image. The intensity-based detector TrackPy drops from $0.73$ to $0.61$ AUC score for localisation within $210nm$, whereas FCNN only drops from an AUC score of $0.99$ to $0.98$ for localisation within $210nm$. 

\subsubsection{Experiment 3 - $1.75\;\mu m^2/s$}
% need 2000 images with diffusion 1.75 mu m^2/s
At last, we make the high-intensity pixels diverge from the microscope PSF as much as possible for a realistic diffusion coefficient value. We use the diffusion coefficient of $1.75 \mu m^2/s$, i.e., ten times larger than the diffusion coefficient used in experiment one. \textbf{\Cref{fig:testdata} (d)} illustrates how increasing motion blur scatters the high-intensity pixels across a larger area and, in some cases, makes the high-intensity pixels interchangeable for noise. Again we see in \textbf{\cref{tab:auc}, column three}, that TrackPy drops in AUC performance, going from $0.61$ to $0.47$, whereas FCNN keeps an AUC score above $0.90$ for localisation within $210nm$.

% Write about the details of the network (very briefly architecture). For each configuration, epoch etc...
% Present the results of the network
% [Rasmus-Nikolay] Illustrations with nice images for detection (maybe something about trackpy ?)

\section{Discussion}
To tackle the problem of accurate localization of QD in pixel-by-pixel recorded images, we developed a simulator that generates a particle undergoing a Brownian motion recorded by a two-photon laser scanning microscope. Implementing the motion of the laser beam during pixel-by-pixel image scanning allows the reproduction of complex motion blur typically observed in experimental data. Note, that our simulator simplifies local transport in vivo and 2PM. First, we compute the trajectory of the QD by Brownian motions and second, we only generate images with one QD showing. To simplify the simulation, we neglected the excess and readout noises \cite{Mortensen2010} introduced by the detector (photomultiplier tube; PMT) and assumed that the actual number of recorded photons is Poisson-distributed. 
We show that realistic simulated data, which incorporates the effect of motion blur, non-specific background fluorescence, and Poisson shot noise, improves the deep learning-based localization of QDs, seen in experiments 1 to 3. The primary result of our experiments demonstrates that the FCNN is robust against motion blur and keeps an AUC score above $0.90$, whereas TrackPy drops the AUC score to $0.47$. 
But we only trained one model. So for future implementations, we intend to make a five-fold cross-validation. 
To our knowledge, this is the first simulator of training data for 2PM. 
It opens for better and faster trajectory recovery and analysis, an invaluable tool to understand how molecules move in complex, turbid, biological tissues, e.g., in the brain. 

\section{Methods}
\subsection{Data description}
We simulated a training dataset of 29838 images and three testing datasets of 2570, 2410 and 1870 images. We chose a training dataset size of 29838 to avoid overfitting, and we chose a test dataset size of 6850 because it is common practice to choose a test dataset that is $20\%$ of the size of the training dataset.  
The following parameters reproduce in vivo experiment parameters: All images were 24 pixels tall and 25 pixels wide, trajectories of the QDs were simulated with a set of diffusion coefficients between $0.75 \mu m^2/s$ and $1.75 \mu m^2/s$, the pixel size was $0.07 \mu m$, the pixel exposure time was $2 \mu s$ and the retracing time was $1.1ms$. 
To ensure single-molecule localisation images contained 1 QD.  

\subsection{Deep Learning Model}
We use the open-access, and easy-to-use PyTorch implementation of FRCNN-keypoint with a resNet backbone \cite{resnet} pre-trained on the MS COCO dataset \cite{coco}.

\subsection{Train Configuration}
We only trained our model using 20 epochs because the model converged. Of the 20 models, we chose the model with the highest validation accuracy. Due to limited GPU memory, we chose a batch size of 6 images, a validation batch size of 4 images, and stochastic gradient descent. We tuned the learning rate to 0.001 \cite{pytorch-vision}. 

\begin{figure}[h!]
    \centering
        \includegraphics[width=\textwidth]{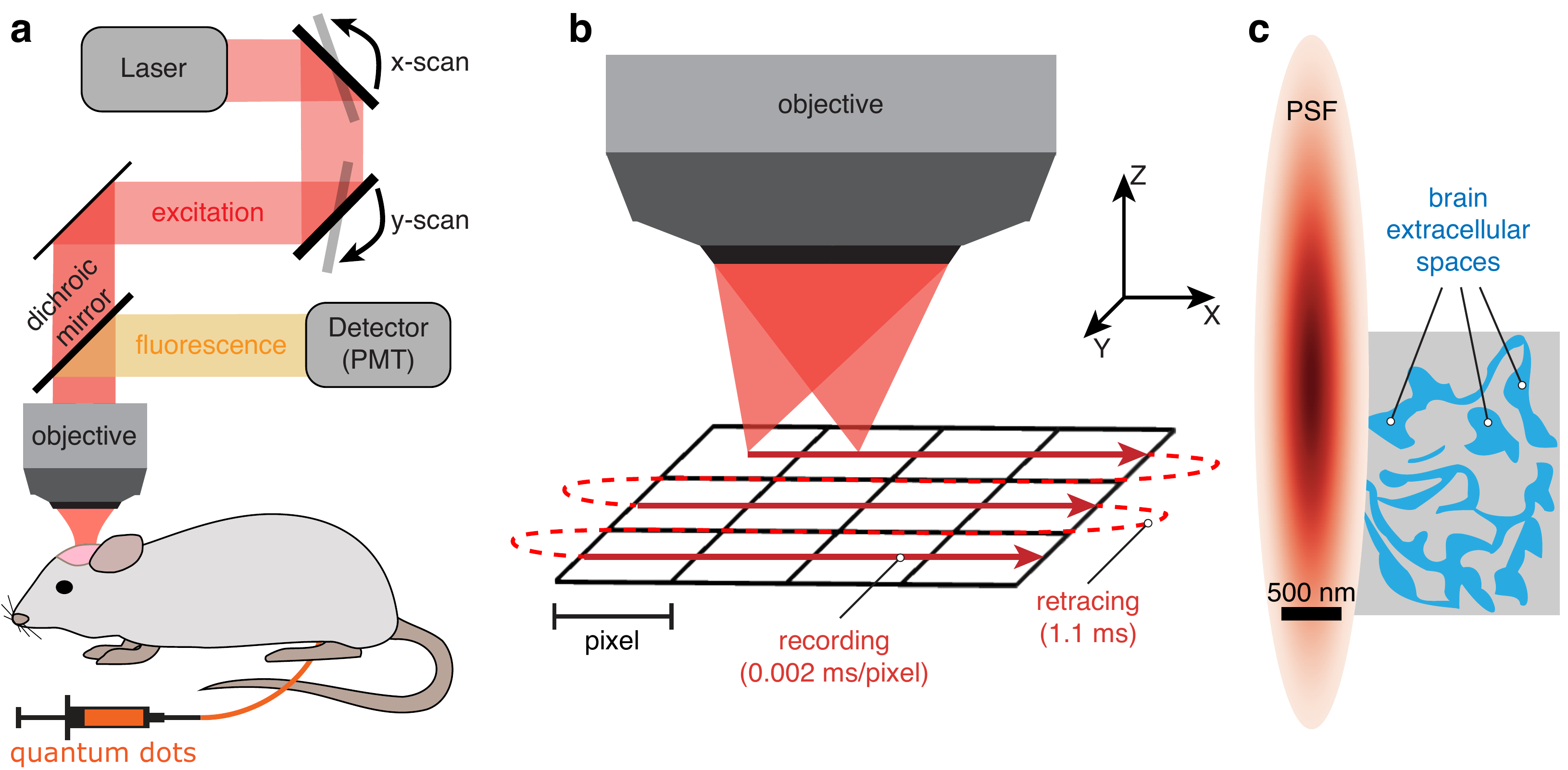} 
    \caption{\textbf{Imaging single quantum dots in the brain of living mice with two-photon microscopy (2PM).}
    \textbf{a} A simplified scheme of the experimental setup: 2PM images quantum dots (QDs) inside the brain of a living, anaesthetised mouse by sweeping the laser beam across the brain and collecting fluorescence emitted by the QDs. 
    \textbf{b} Images are recorded by moving the laser beam (by rotating x- and y-scan mirrors shown in \textbf{a}) along straight lines (solid red). 
    Between the lines, the beam is retraced (dashed red): this ensures the same scanning speed as the laser beam. 
    \textbf{c} Dimensions of the point spread function (PSF; drawn to scale) of a 2PM inside a brain \cite{Kutuzov2018} and a scheme of brain extracellular spaces --- a 3D network of channels between brain cells membranes --- where QDs are moving. 
    In 2PM, the PSF represents the square of the laser intensity near the focus and it is approximately Gaussian, with standard deviation along the z-axis being $\approx$6 times larger than along x- and y-axes. }
    \label{fig:2pm}
\end{figure}

\subsection{Materials}

We used a commercial two-photon scanning microscope (FVMPE-RS) built by Olympus (Japan), equipped with a Ti:Sapphire laser and GaAsP detectors (photomultiplier tubes; PMTs).
We used commercial QTracker655 quantum dots (QDs; ThermoFisher Scientific, USA), which we diluted 15 times in saline before intravenous injection into anaesthetized mice.
%For model experiments, we dissolved QDs in 54\% sugar syrups.
%We chose this concentration (concentration affects viscosity, hence, diffusion coefficient of the QDs) because it allowed us to reproduce the motion blur observed on images recorded in the brain.

\subsection{Animals and surgical procedures}
All protocols involving mice were approved by the Danish National Committee on Health Research Ethics, in accord with the guidelines of the European Council’s Convention for the Protection of Vertebrate Animals used for Experimental and Other Scientific Purposes, and complied with the ARRIVE guidelines.
We used four male wild type mice (C57bl/6j; Janvier-labs).
We followed standard surgical procedures including tracheotomy, cauterization, and craniotomy, described in detail elsewhere \cite{Kucharz2021}.
QDs were injected intravenously and released in the the brain parenchyma by ablating a small capillary with high laser power.

\subsection{Simulating 2PM images of diffusing QDs.}
\subsubsection{Laser beam motion simulation} First, we calculated the trajectory of the laser beam during image acquisition (Fig.~\ref{fig:2pm}\textbf{b}) assuming that the laser beam moves along straight-line segments at the constant speed of 2 $\mu m$ per pixel (typically used in in vivo experiments) until it reaches the end of a scanning line.
Then the laser beam is retraced (no photons are recorded during this phase) to the beginning of the next line, which takes 1.1 ms. 
The process is repeated until the whole image is recorded.
We estimated the position of the laser beam with the same time step as for diffusion trajectories ($\delta t$).
\subsubsection{Random walk} Second, we simulated a random walk trajectory (independently for all three coordinates, $xyz$) of QDs during image acquisition by cumulatively adding normally-distributed random numbers $\mathcal{N}(0,\sigma^2)$, with zero mean and variance $\sigma^2 = 2D\delta t$, where $D$ is the diffusion coefficient of the QDs and $\delta$ is a time step.
To align the timestep of the simulator with the timestep $\delta$, we needed to sample from two normal distributions. The first distribution uses a timestep of $2\mu m$, whereas the second uses the retracing timestep of $1.1 ms$.
% We used a diffusion coefficient $D = 1\,\mu$m$^2$/s, which falls in between experimentally measured diffusion coefficients of a 70-kDa fluorescent dextran and a QD (larger than the QDs we used) in the brain in vivo \cite{Thorne2006}. 
%
\subsubsection{Point Spread Function} Third, we modelled the expected photon count in individual image pixels. 
Assuming that the square of the laser light intensity (two photons are absorbed simultaneously at 2PM) near the focus is approximately a 3D Gaussian \cite{MPM} with standard deviations $\sigma_{xy}$ in the $xy$-plane and $\sigma_z$ (depth) along the optical axis (Fig.~\ref{fig:2pm}\textbf{c}), the expected number of photons is 
\begin{align}
    N=\frac{N_{qd}}{2\pi\sigma_{xy}^2}\exp{\left(-\frac{(X - x)^2 + (Y - y)^2}{2\sigma_{xy}^2} \right)} \cdot \frac{1}{\sqrt{2\pi\sigma_z^2}}\exp{\left(-\frac{(Z - z)^2}{2\sigma_z^2}\right)} \label{eq:psf} + N_{bkg}
\end{align}
with $(X, Y, Z)$ the position of the QD, $(x,y,z)$ the position of a laser beam, $N_{qd}$ is the total possible number of emitted photons by a QD in the focal plane in the direction of the objective, and $N_{bkg}$ is expected number of photons of the fluorescent background (autofluorescence). 
We used the value $N_{bkg} = 8.4$, which was estimated by averaging the background fluorescence on images recorded in the brain.
The value of $N_{qd}$ was adjusted to imitate the signal-to-noise ratio of the brain images.
We used realistic, measured in vivo, parameters of the 2PM PSF: $\sigma_{xy} = 0.21$ $\mu$m and $\sigma_{z} = 1.36$ $\mu$m \cite{Kutuzov2018}.

%Finally, we assumed that the actual number of recorded photons is Poisson-distributed with the expected value $N$ defined above.
%Note that we neglected the excess and readout noises \cite{Mortensen2010} introduced by the detector (photomultiplier tube; PMT) to simplify the simulation.
%If needed, more realistic distributions of detector output can be used.

\subsection{Localization of QDs}
Several DL-based detectors have been proposed in recent years \cite{redmon-etal:2015,lin-etal:2017}, summarised in a recent review \cite{Youzi-etal:2020}. 
Many of these detectors have been tailored to detect complex objects. QDs are simple practically point sources of light but they do not image as the microscope's PSF because of the motion blur. Furthermore, images of QDs can be very noisy due to the high fluorescence background or movement of the QD out of focus. For such problems, the Faster R-CNN-Keypoint (FRCNN) \cite{maskfcnn,pytorch-keypoint-rcnn, pytorch}, which predicts object bounding boxes along with key points, has proven very reliable and has available implementations and was adopted in this work. 

\subsection{Metrics}
 For training the model we have used classical detection metrics for the bounding boxes and keypoints predicted by FRCNN. Detection speed in frames per second is of no interest in our problem. Precision and recall \cite{deepLearningSurvey} are based on intersection over union (IoU):
\begin{align}
    IoU(\hat{B}, B) = \frac{area(\hat{B} \cap B)}{area(\hat{B} \cup B)} \label{eq:IoU}
\end{align}
with $\hat{B}$ is a predicted bounding box and $B$ the associated ground truth box.
Determining true positives (TP), false positives (FP), true negatives (TN), false negatives (FN) for both bounding box and keypoints are computed according to the COCO standard \cite{coco,cocoapi}.
For evaluation, we needed another metric since an intensity-based detector such as TrackPy does not predict bounding boxes \cite{trackpy}. Therefore, we computed the area under the curve (AUC) from the precision and the recall. In addition, we modified the precision and the recall seen in \cite{objectDetectionBasedOnDeepLearning}  to use an L2 distance instead of an IoU constraint. 
\begin{align}
    \textit{precision} &= \frac{tp}{tp+fp} \label{eq:precision}\\
    \textit{recall} &= \frac{tp}{tp+fn} \label{eq:recall}
\end{align}
That means we define a prediction to TP if the distance between the predicted key point and the ground truth key point is below or equal to a threshold $\varepsilon$. FP is defined as if the distance exceeds the threshold or the prediction is duplicated. At last, FN is defined as if the detector fails to predict. We then computed the precision and the recall using equation \cref{eq:precision,eq:recall}. Finally, we computed the area under the curve (AUC) using precision and recall. Precision tells the probability that a prediction is correct, whereas recall tells the probability of detecting all possible ground truths. The AUC combines the metric, e.g. If the AUC score is close to 1, the detector correctly predicts and detects all possible ground truths. If the AUC score is close to 0, the detector either poorly predicts, fails to predict or both.

\section*{Declarations}
\subsection*{Funding}
Not applicable 
\subsection*{Conflict of interest/Competing interests}
Not conflict or competing interests
\subsection*{Ethics approval}
Not applicable
\subsection*{Consent to participate}
Not applicable
\subsection*{Consent for publication}
Not applicable
\subsection*{Availability of data and materials}
We released the data as open-access at the time of publication. 
\subsection*{Code availability}
Most of the code in this project is custom-made and will be open-access at the time of publication. However, we are happy to share the code upon request. 
\subsection*{Authors' contributions}
The first and second co-first authors contributed equally to this work. They made the primary work of this paper. The third, fifth and sixth authors contributed with writing, supervision and concept development. The fourth author helped with the programming  


\bibliography{sn-bibliography, mybibliography, refs1}% common bib file


\end{document}